\documentclass[aps,prl,showpacs,showkeys,amsmath,amssymb,twocolumn,floatfix]{revtex4-1}
\usepackage{graphicx}
\usepackage{color}
\usepackage{enumitem}
\usepackage{bm}

\newcommand{\nubar}[1]{$\overline{\nu}_{#1}$}

\begin{document}

\title{Spectral Structure of Electron Antineutrinos from Nuclear
  Reactors}

\author{D.~A.~Dwyer}
\email{dadwyer@lbl.gov}
\affiliation{Lawrence~Berkeley~National~Laboratory, Berkeley, CA, USA}
\author{T.~J.~Langford}
\email{thomas.langford@yale.edu}
\affiliation{Yale~University, New~Haven, CT, USA}
\date{\today}					    

\begin{abstract}

Recent measurements of the positron energy spectrum obtained from
inverse beta decay interactions of reactor electron antineutrinos show
an excess in the 4 to 6~MeV region relative to current predictions.
First-principle calculations of fission and beta decay processes
within a typical pressurized water reactor core identify prominent
fission daughter isotopes as a possible origin for this excess.  These
calculations also predict percent-level substructure in the
antineutrino spectrum due to Coulomb effects in beta decay.  Precise
measurement of this substructure can constrain nuclear reactor
physics.  The substructure can be a systematic uncertainty for
measurements utilizing the detailed spectral shape.

\end{abstract}

\maketitle

\section{Introduction}

Determination of the mixing angle $\theta_{13}$ required a new
generation of reactor antineutrino experiments with unprecedented
statistical precision~\cite{Abe:2011fz, An:2012eh, Ahn:2012nd}.  The
Daya Bay and RENO experiments have each detected $\sim$10$^{6}$
reactor \nubar{e} interactions~\cite{An:2013zwz, fortheRENO:2013lwa}.
Proper characterization of the \nubar{e} energy spectrum emitted by
nuclear reactors is important for such measurements of neutrino
properties.  The standard method of modeling the \nubar{e} emission by
nuclear reactors relies on the correlation between the energy spectra
of the $\beta^-$ and \nubar{e} in beta decay.  Here we refer to this
method as $\beta^-$ {\em conversion}.  For a single beta decay, the
prediction of the \nubar{e} spectrum from the measured $\beta^-$
spectrum can be done with high precision.  In the 1980's, foils of the
fissile isotopes $^{235}$U, $^{239}$Pu, and $^{241}$Pu were exposed to
a thermal neutron flux from the ILL reactor, and the cumulative beta
decay $\beta^-$ spectra of the fission daughters were
measured~\cite{Schreckenbach:1985ep, VonFeilitzsch:1982jw,
  Hahn:1989zr}.  More recently, a similar measurement was made for
$^{238}$U~\cite{Haag:2013raa}.  The fission of four main parent
isotopes represent $>$99\% of reactor \nubar{e} emission.  Given that
each measured $\beta^-$ spectrum is composed of thousands of unique
beta decays, the conversion must be done en masse.  This introduces
uncertainties of a few percent in the corresponding prediction of the
cumulative \nubar{e} spectra.  Detailed descriptions of such
calculations can be found in~\cite{Carter:1959qm, Mueller:2011nm,
  Huber:2011wv}.  A recent study suggested that the uncertainties in
conversion of the $\beta^-$ to \nubar{e} spectrum may have been
underestimated due to shape corrections for forbidden beta
decays~\cite{Hayes:2013wra}.

In this note we discuss an alternative calculation of \nubar{e}
emission by nuclear reactors based on nuclear databases.  This {\em ab
  initio} approach relies on direct estimation of the \nubar{e}
spectrum from the existing surveys of nuclear data.  This method
suffers from rather large uncertainties in our knowledge of the
fission and decay of the $>$1000 isotopes predicted to be present in a
nuclear reactor core.  Despite these uncertainties, the {\em ab
  initio} calculation predicts a spectral bump with
$E_{\overline{\nu}}$=5--7~MeV ($E_{e+}$=4--6~MeV) relative to the
$\beta^-$ {\em conversion} method.  Recent measurements of the
positron energy spectra from \nubar{e} inverse beta decay
($\overline{\nu}_e + p \rightarrow e^{+} + n$) show a similar
$\sim$10\% excess of positrons detected with energies from 4 to 6~MeV.
We also observe substructure at the level of a few percent in the
calculated energy spectra, which is difficult to demonstrate from the
$\beta^-$ {\em conversion} method.  This substructure is due to
discontinuities introduced by the Coulomb phase space correction in
the \nubar{e} spectrum of each unique decay branch.  Precise
measurement of this substructure could provide a unique handle on the
nuclear physics occurring within a reactor.  When not predicted in the
model, the substructure may present a systematic uncertainty for
measurements relying on high-resolution features of the reactor
\nubar{e} energy spectrum, for example~\cite{Learned:2006wy,
  Li:2013zyd}.

\section{Calculation of the \nubar{e} Spectrum}

The {\em ab initio} method of calculating the \nubar{e} spectrum
follows that presented in~\cite{Vogel:1980bk, Hayes:2013wra}.
Considering a reactor operating at equilibrium, the total antineutrino
spectrum can be estimated as the sum of a large number of beta decay
spectra,
\begin{equation}\label{eq:totalSpectrum}
  S(E_{\overline{\nu}}) = \sum\limits_{i=0}^{n}R_i\sum\limits_{j=0}^{m}f_{ij}S_{ij}(E_{\overline{\nu}}).
\end{equation}
The equilibrium decay rate of isotope $i$ in the reactor core is
$R_i$.  The isotope decays to a particular energy level $j$ of the
daughter isotpe with a relative probability, or branching fraction,
$f_{ij}$.  The antineutrino spectrum for each decay branch is given by
$S_{ij}(E_{\overline{\nu}})$.  The collective \nubar{e} emission from
a reactor is due to $>$1000 daughter isotopes with $>$6000 unique beta
decays.

Estimation of the decay rates $R_i$ depend on our knowledge of the
nuclear processes within the reactor core.  For a fission of a parent
nucleus, $^{A}_{Z}N_p$, the propability of fragmenting to a particular
daughter nucleus $^{A'}_{Z'}N_d$ is given by the {\em instantaneous}
yield, $Y^{i}_{pd}$.  The majority of these fission daughters are
unstable, and will decay until reaching a stable isotopic state.  The
{\em cumulative} yield $Y^{c}_{pi}$ is the probability that a
particular isotope $^{A'}_{Z'}N_i$ is produced via the decay chain of
any initial fission daughter.  On average, the daughter isotopes of
each fission undergo 6 beta decays until reaching stability.  For
short-lived isotopes, the decay rate $R_i$ is approximately equal to
the fission rate $R^f_p$ of the parent isotope $p$ times the
cumulative yield of the isotope $i$,
\begin{equation}
  R_i \simeq \sum\limits^{P}_{p=0}R^{f}_{p} Y^{c}_{pi}
\end{equation}
The ENDF/B.VII.1 compiled nuclear data contains tables of the
cumulative fission yields of 1325 fission daughter isotopes, including
relevant nuclear isomers~\cite{England:1992aa, Chadwick:2011}.
Evaluated nuclear structure data files (ENSDF) provide tables of known
beta decay endpoint energies and branching fractions for many
isotopes~\cite{Tuli:1996jt}.  Over 4000 beta decay branches are found
which have endpoints above the 1.8~MeV threshold for inverse beta
decay.  The spectrum of each beta decay $S_{ij}(E_{\overline{\nu}})$
was calculated including Coulomb~\cite{Schenter:1983aa},
radiative~\cite{Sirlin:2011wg}, finite nuclear size, and weak
magnetism corrections~\cite{Hayes:2013wra}. In the following
calculations we begin by assuming that all decays have allowed
Gamow-Teller spectral shapes.  The impact of forbidden shape
corrections will be discussed later in the text.

\begin{figure}[htb]
\includegraphics[width=0.95\columnwidth]{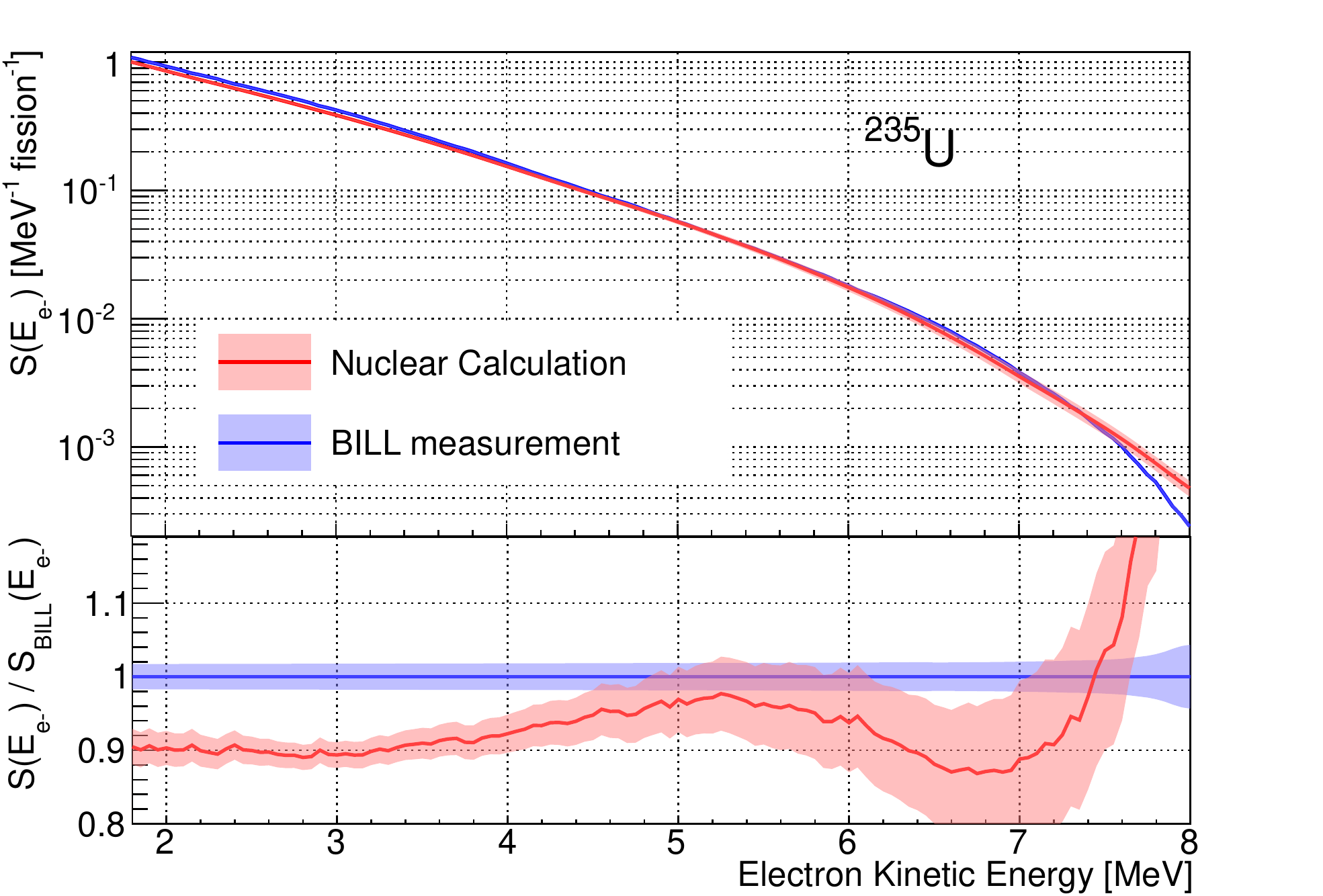}
\includegraphics[width=0.95\columnwidth]{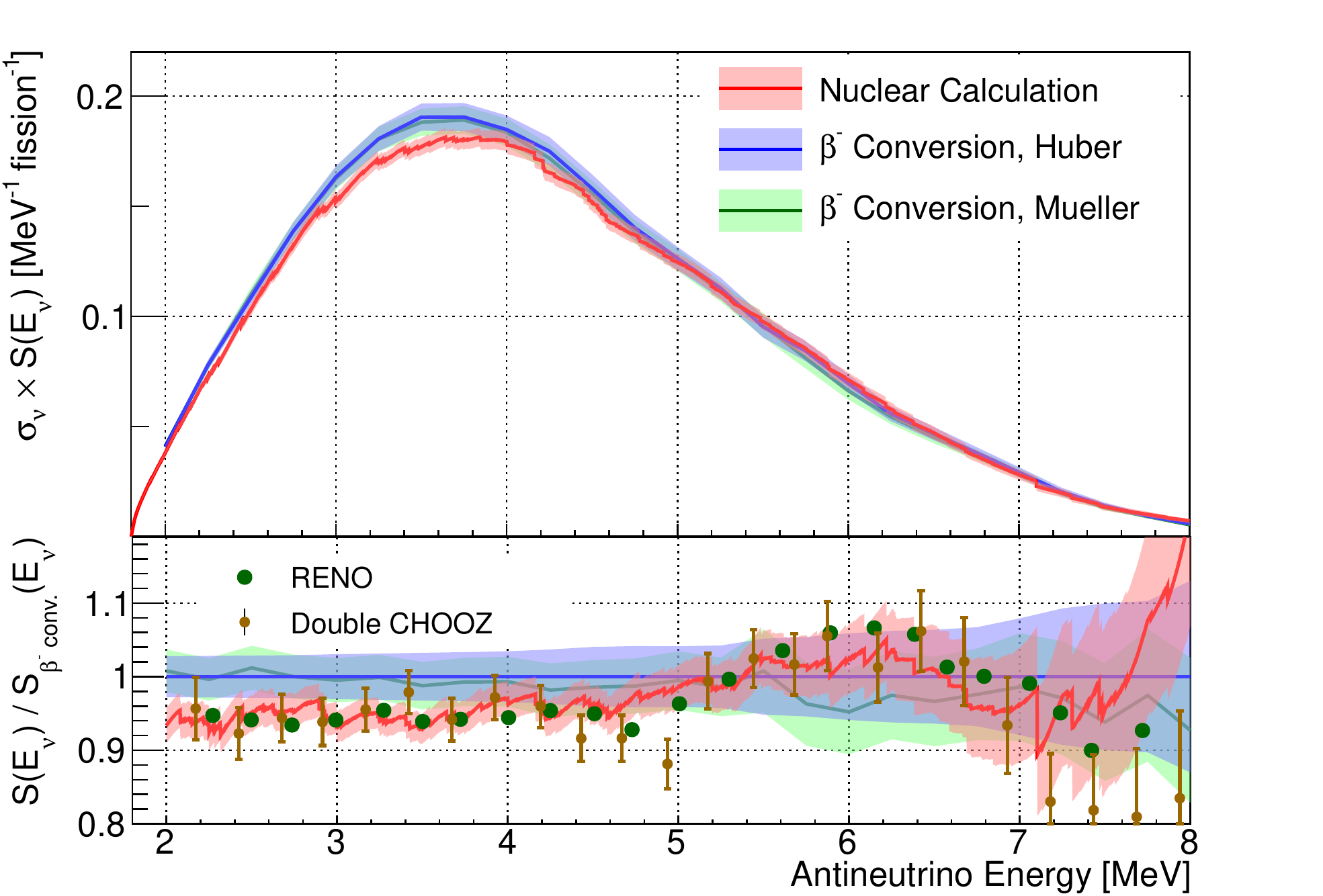}
\caption{ Upper: The {\em ab initio} nuclear calculation of the
  cumulative $\beta^-$ energy spectrum per fission of $^{235}$U
  exposed to thermal neutrons (red), including 1-$\sigma$
  uncertainties due to fission yields and branching fractions.  The
  measured $\beta^-$ spectrum from~\cite{Schreckenbach:1985ep} is
  included for reference (blue).  Lower: The corresponding \nubar{e}
  spectrum per fission in a nominal reactor weighted by the inverse
  beta decay cross section (red), compared with that obtained by the
  $\beta^-$ {\em conversion} method (blue~\cite{Huber:2011wv},
  green~\cite{Mueller:2011nm}).  See text for discussion of
  uncertainties.  Measurements of the positron spectra
  (green~\cite{renoNu2014}, brown~\cite{Abe:2014bwa}) are similar to
  the {\em ab initio} calculation, assuming the approximate relation
  $E_{\overline{\nu}}\simeq E_{e^+}+0.8$~MeV.
\label{fig:u235Calc}}
\end{figure}

The upper panel of Fig.~\ref{fig:u235Calc} shows the electron spectrum
per fission of $^{235}$U calculated according to
Eq.~\ref{eq:totalSpectrum}.  The $\beta^-$ spectrum measured in the
1980s using the BILL spectrometer is shown for
comparison~\cite{Schreckenbach:1985ep}.  Both spectra are absolutely
normalized in units of electrons per MeV per fission.  The lower panel
shows the calculated \nubar{e} spectrum for a nominal nuclear reactor
with relative fission rates of 0.584, 0.076, 0.29, 0.05 respectively
for the parents $^{235}$U, $^{238}$U, $^{239}$Pu, $^{241}$Pu.  The
spectra have been weighted by the cross section for inverse beta decay
to more closely correspond to the spectra observed by experiments.
Prediction of the \nubar{e} spectrum by $\beta^-$ {\em conversion} of
the BILL measurements~\cite{Mueller:2011nm, Huber:2011wv} shows a
different spectral shape.  In particular, there is a bump near 6~MeV
in the calculated spectrum not shown by the $\beta^-$ {\em conversion}
method.  Note that the hybrid approach of Ref.~\cite{Mueller:2011nm}
used the {\em ab initio} calculation to predict most of the $\beta^-$
and \nubar{e} spectra, but additional fictional $\beta^-$ branches
were added so that the overall electron spectra would match the BILL
measurements.  These corresponding \nubar{e} spectra for these
branches were estimated using the $\beta^-$ {\em conversion} method.
Since this method is constrained to match the BILL measurements, it is
grouped with the other $\beta^-$ {\em conversion} predictions.

The significant differences between the calculation and BILL
measurements are generally attributed to systematic uncertainties in
the {\em ab initio} calculation.  The 1-$\sigma$ uncertainty bands
presented here include only the stated uncertainties in the cumulative
yields and branching fractions.  Three additional systematic
uncertainties are prominent but not included: data missing from
nuclear databases, biased branching fractions, and beta decay spectral
shape corrections.

{\em Missing Data:} It is possible that the ENDF/B tabulated fission
yields lack data on rare and very short lived isotopes in regions far
from nuclear stability.  In~\cite{Vogel:1980bk} it was argued that
this missing data would favor higher-energy decays.  For the known
fission daughters, $\sim$6\% of the yielded isotopes have no measured
beta decay information.  Both of these effects result in an
underprediction of the spectrum at all energies.

{\em Biased Branching Fractions:} The branch information for known
isotopes may be incomplete or biased.  For example the pandemonium
effect can cause a systematic bias enhancing branching fractions at
higher energies relative to those at lower
energies~\cite{Hardy:1977bd}.  Such a bias would cause an
underprediction of the spectrum at low energies and an overprediction
at high energies.

{\em Shape Corrections:} The beta decay spectra of each branch may
vary from the allowed shape depending on the nuclear matrix elements
connecting initial and final states.  In general these corrections are
small for allowed or slightly forbidden decays, but can be more
significant for those decays involving a large ${\Delta}J$ or
cancellations between matrix elements.  In~\cite{Hayes:2013wra} it was
shown that $\sim$25\% of known reactor \nubar{e} decay branches are
forbidden, and that shape corrections could in principle impact the
$\beta^-$ {\em conversion} method.

These systematic uncertainties are difficult to quantify and do hinder
the absolute prediction of the \nubar{e} rate and spectrum from a
reactor.  To correctly model and incorporate all of these
uncertainties requires an extensive study not considered for this
manuscript.  Instead here we focus on two characteristics of the
calculation which appear robust to these uncertainties.  First, the
combined distribution of beta decay branches predicts a bump in the
antineutrino spectrum from 5 to 7~MeV.  Second, the Coulomb
corrections introduce detailed structure to the \nubar{e} spectrum
that is not reflected in the corresponding $\beta^-$ spectrum.

\section{\label{sec:shoulder}Spectral Shape in 5--7 MeV}

\begin{figure}[htb]
\includegraphics[width=0.95\columnwidth]{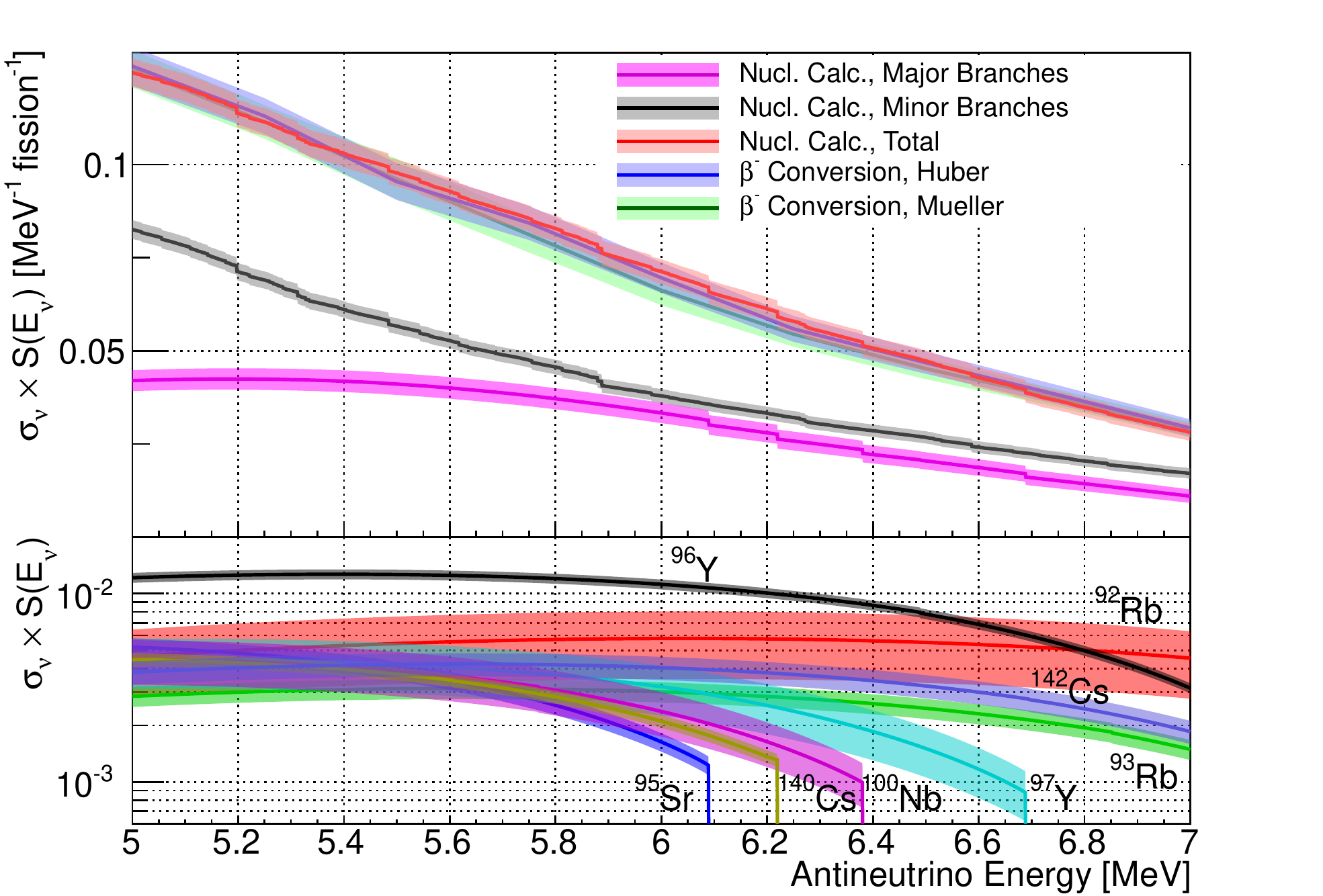}
\caption{ Upper: The calculated reactor \nubar{e} spectrum per fission
  in a nominal nuclear power reactor multiplied by the cross section
  for inverse beta decay (red), in the 5--7~MeV region.  The eight
  most prominent decay branches in this region provide 42\% of the
  total counts (magenta), and combine to produce a bump relative to
  the $\beta^-$ {\em conversion} method (blue~\cite{Huber:2011wv},
  green~\cite{Mueller:2011nm}).  The remaining $>$1100 decay branches
  each provide less than 2\% of the total rate in this region, and
  combined provide a smooth shape (black).  Lower: Individual spectra
  from the eight most prominent branches.  Uncertainties are the same
  as for Fig~\ref{fig:u235Calc}.
\label{fig:shoulderRegion}}
\end{figure}

Recent measurements present a $\sim$10\% excess in the positron
spectrum from inverse beta decay in the region of $E_{e+}$=4--6~MeV
($E_{\overline{\nu}}$=5--7~MeV), similar to the {\em ab initio}
calculation.  In this region, the spectral shape is dominated by eight
prominent decay branches which contribute 42\% of the calculated
rate. All eight branches are transitions between the ground states of
the initial and final isotopes, and all are first forbidden non-unique
decays.  The remaining $\sim$1100 decay branches each contribute at
most 2\% of the total rate, and individually have little influence on
the spectral shape.  Fig.~\ref{fig:shoulderRegion} shows the {\em ab
  initio} prediction broken into the eight major branches and the
remaining minor branches.  Table~\ref{tab:majorBranches} summarizes
these prominent decay branches.

\begin{table}[htb]
\begin{tabular}{|l|r|r|r|c|r|r|}
\hline
Isotope & Q[MeV] & $t_{1/2}$[s] & log($ft$) & Decay Type & $N$[\%] & $\sigma_N$[\%] \\ \hline
 $^{96}$Y & 7.103 & 5.34 & 5.59 & $0^-\rightarrow0^+$ & 13.6 & 0.8  \\ \hline
 $^{92}$Rb & 8.095 & 4.48 & 5.75 & $0^-\rightarrow0^+$ & 7.4 & 2.9 \\ \hline
 $^{142}$Cs & 7.308 & 1.68 & 5.59  & $0^-\rightarrow0^+$ & 5.0 & 0.7 \\ \hline
 $^{97}$Y & 6.689 & 3.75 & 5.70 & $1/2^-\rightarrow1/2^+$ & 3.8 & 1.1 \\ \hline
 $^{93}$Rb & 7.466 & 5.84 & 6.14 & $5/2^-\rightarrow5/2^+$ & 3.7 & 0.5 \\ \hline
 $^{100}$Nb & 6.381 & 1.5 & 5.1 & $1^+\rightarrow0^-$ & 3.0 & 0.8 \\ \hline
 $^{140}$Cs & 6.220 & 63.7 & 7.05 & $1^-\rightarrow0^+$ & 2.7 & 0.2 \\ \hline
 $^{95}$Sr & 6.090 & 23.9 & 6.16 & $1/2^+\rightarrow1/2^-$ & 2.6 & 0.3 \\ \hline
\end{tabular}
\caption{Most prominent beta decay branches in the region of
  $E_{\overline{\nu}}$=5--7~MeV. The table presents the decay parent,
  endpoint energy, half-life, and decay $ft$ value.  The decay type
  describes the parent and daughter states.  The moderate $ft$ values
  and lack of significant change of $J^\pi$ suggest that all but
  possibly $^{140}$Cs decay with allowed spectral shapes.  The rate
  each branch contributes to the total between 5--7~MeV is $N$,
  accounting for the inverse beta decay cross section.  The 1-$\sigma$
  uncertainty due to the fission yield and branching fraction is
  $\sigma_{N}$.\label{tab:majorBranches}}
\end{table}

The impact of each unquantified systematic uncertainty on this
spectral feature can be examined.  Contributions from missing nuclear
data could add additional decay branches in this region, increasing
the overall normalization and difference from the $\beta^-$ {\em
  conversion} model.  To remove the bump-like shape, it would require
that the additional branches have a particular distribution of
endpoints just below and just above the excess.  While possible, this
seems contrived.  For the eight prominent branches, six are 0$^-$
decays.  These decays are not expected to have any significant
deviation from allowed shapes.  The $ft$ values are mostly in the 5 to
6 range, consistent with allowed shapes.  Only $^{140}$Cs has a large
$ft$ value and decay type consistent with a possible forbidden shape
correction. Since this isotope contributes only 2.7\% of the rate, the
resulting correction should be small.  Forbidden shape corrections on
the numerous minor branches can only negligibly impact the overall
structure, although a cumulative effect could slightly impact the
normalization and slope.  The current uncertainty band includes the
stated uncertainties on the branching fractions.  Biases such as the
pandemonium effect would need to be significantly larger than these
uncertainties on the eight major branches in order to remove the
bump-like shape.  Pandemonium corrections on the large number of minor
branches could slightly reduce the total normalization and change the
slope in this region.  In particular, $^{92}$Rb suffers from
significant uncertainty in the branching fraction to the ground state.
Our calculation used a branching fraction of 51$\pm$18\%
from~\cite{Baglin:2000bg}.  In~\cite{Baglin:2012bg} it was changed to
95$\pm$0.7\% to correct for a corresponding overestimation of
branching fractions for known excited levels.  Recent measurements
suggest this may actually be due to unknown excited levels, providing
a preliminary result of 74\%~\cite{Zakari:2014az}.  Awaiting a
definitive measurement, we retain the older value and larger
uncertainty.  While these uncertainties could reasonably impact the
normalization and slope of the spectrum in this region, the prediction
of a bumped shape seems robust provided the tabulated data for these
prominent branches is correct.

Fig.~\ref{fig:u235Calc} includes the recently measured deviations in
the positron spectrum from inverse beta decay~\cite{renoNu2014,
  Abe:2014bwa}.  The relation $E_{\overline{\nu}}\simeq
E_{e^+}+0.8$~MeV was used to approximate the ratio for \nubar{e}
energy.  Normalization was adjusted to provide a comparison of only
spectral shape.  Each experimental spectrum includes percent-level
systematic effects from detector resolution and nonlinearity not
present in the calculation, providing only an illustrative comparison.
Given these assumptions and the model uncertainties already discussed,
the overall agreement between the data and {\em ab initio} calculation
is surprising.

\section{\label{sec:fineStructure}Detailed Spectral Substructure}

\begin{figure}[htb]
\includegraphics[width=0.95\columnwidth]{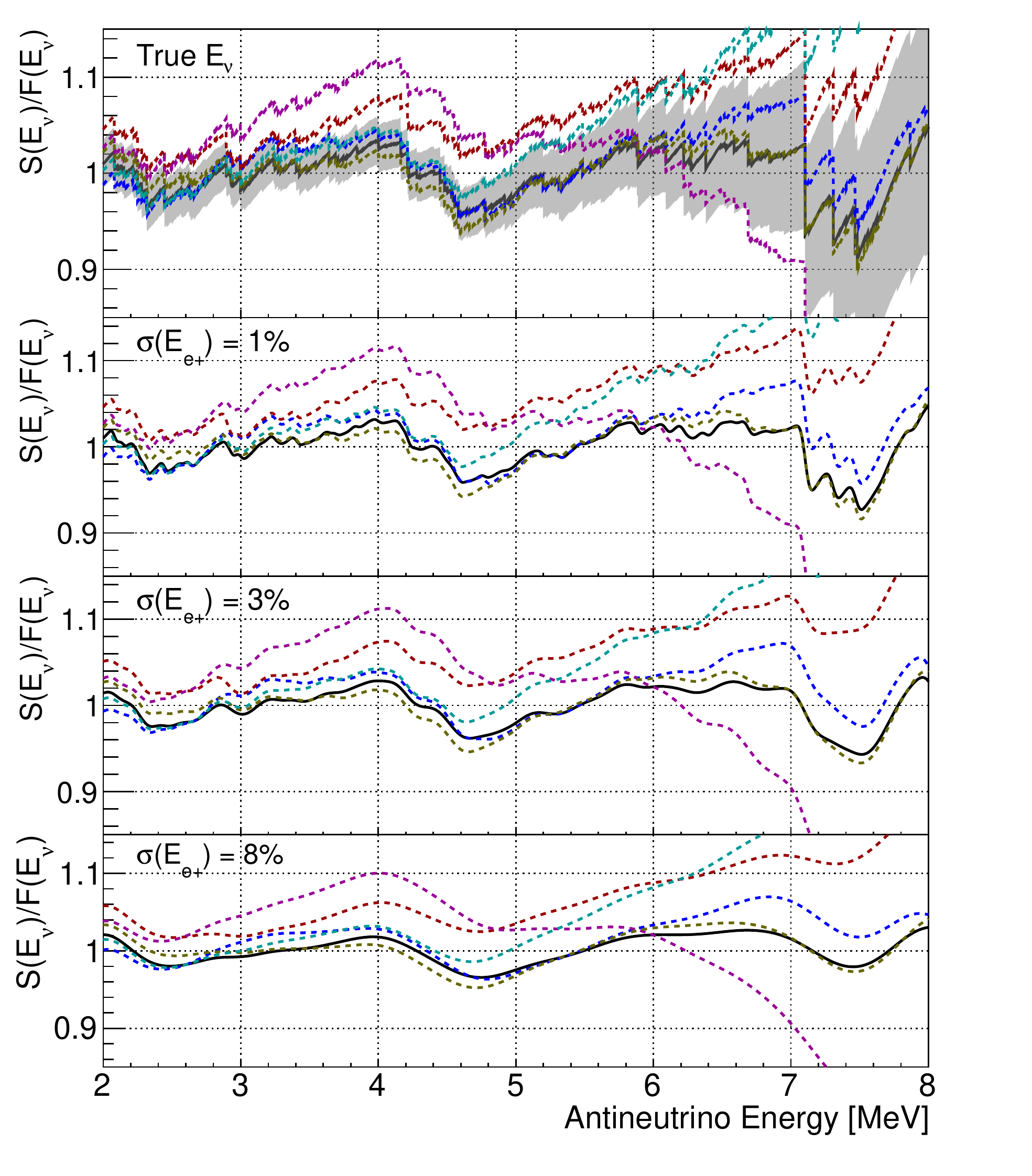}
\caption{ Upper panel: The calculated \nubar{e} energy spectrum from a
  nominal nuclear reactor (black line) divided by a smooth
  approximation [Eq.~(\ref{eq:smoothSpectrum})], including the
  1-$\sigma$ uncertainties due to the fission yields and branching
  fractions (grey band).  Significant discontinuities are caused by
  the Coulomb correction to the spectra of prominent beta decays.
  Random variation of fission yields and branching fractions can alter
  the particular pattern (colored lines).  Lower panels: The same
  spectra after accounting for detector resolution.  The current
  generation of experiments with $\sim$8\% resolution are sensitive to
  the larger variations.  Future high-resolution experiments would
  detect significant substructure.
\label{fig:substructure}}
\end{figure}

The {\em ab initio} calculated spectrum shows detailed substructure
due to beta decay Coulomb corrections.  The effect of the Coulomb
correction on a single decay branch can be seen as the sharp
discontinuity at the endpoint of the \nubar{e} spectra.  There is no
corresponding detailed structure in the $\beta^-$ spectrum since the
Coulomb corrections impact the low-energy end of the electron
spectrum.  The substructure in the {\em ab initio} calculation is most
apparent in the ratio relative to a smooth analytic
approximation~\cite{PhysRevD.39.3378},
\begin{equation}\label{eq:smoothSpectrum}
F(E_{\overline{\nu}})={\rm exp}(\sum_i\alpha_iE_{\overline{\nu}}^{i-1}).
\end{equation}
A fit to the calculated spectrum provides $\bm{\alpha}=\{0.4739,
0.3877, -0.3619, 0.04972, -0.002991\}$.  A significant number of
discontinuities are present with amplitudes of a few percent or
greater, as shown in Fig.~\ref{fig:substructure}.

Systematic uncertainties in the {\em ab initio} calculation introduce
variation in the specific pattern of this substructure.  In
alternative calculations, random gaussian fluctuations were applied to
the yields and branching fractions according to the tabulated
1-$\sigma$ uncertainties.  Parameters were not allowed to fluctuate to
negative values, introducing a bias toward enhancing the overall
spectrum.  Fig.~\ref{fig:substructure} shows five example spectra from
these calculations.

The unquantified systematic uncertainties can also modify the
substructure.  Missing nuclear data would introduce additional
isotopes and decay branches, thereby increasing the number of
discontinuities.  The pandemonium effect would slightly reduce the
amplitude of discontinuities at high energies, and enhance those at
low energies.  Shape corrections can increase or decrease the
amplitude of a particular discontinuity.  While these uncertainties
can make the exact pattern of substructure difficult to predict, it is
clear that substructure of the scale shown will be present.

Current reactor \nubar{e} experiments have sufficient resolution (6 to
8\%$\times\sqrt{E_{\rm e+}/{\rm 1~MeV}}$) and statistical precision to
be sensitive to these detailed spectral features.
Fig.~\ref{fig:substructure} demonstrates the spectral structure after
accounting for detector resolution in the measurement of positrons
from inverse beta decay.  Measurements reaching percent-level
resolution would reveal significant details of the nuclear processes
occuring within a reactor.  Once measured, the structure may also
prove useful for calibration of future detectors.  With further study
it could possibly serve as a diagnostic for nuclear reactor operation.
These features can pose an additional systematic uncertainty in
measurements relying on the spectral shape.  For example, proposed
measurements of the neutrino mass hierarchy using reactor \nubar{e}
require detectors with an energy resolution of at least
3\%~\cite{Learned:2006wy, Li:2013zyd}.  Given that the mass hierarchy
presents itself as small differences in the high-frequency oscillatory
pattern in the spectrum, the detailed spectral structure may
complicate the measurement.

\section{Discussion}

While there are still significant uncertainties in {\em ab initio}
calculation of the reactor \nubar{e} energy spectrum, two specific
characteristics are predicted.  A spectral bump due to prominent beta
decay branches in the 5--7~MeV region is similar to that seen in
recent measurements.  Dedicated studies of the fission yields and
branching fractions of these prominent decays would help confirm the
spectral shape in this region.  The presence of this bump in both the
calculated electron and antineutrino spectra suggests that the
discrepancy may not be due to systematics of the $\beta^-$ {\em
  conversion} method, but instead may be an artifact of the original
$\beta^-$ measurements.

Calculation can also predict the level of substructure in the
\nubar{e} energy spectrum from a reactor, but not the particular
pattern.  This may pose an additional challenge for measurements
probing high-resolution features in the spectrum.  Conversely, a
high-resolution measurement of the reactor antineutrino spectrum could
provide useful information for the modeling of nuclear fission within
the reactor core.

These conclusions demonstrate the value of precise measurement of the
\nubar{e} energy spectra from nuclear reactors, reinforcing the
conclusion of Ref.~\cite{Hayes:2013wra}.  Research reactors could
provide a model system (primarily $^{235}$U) for comparison of
measurement and calculation.

\begin{acknowledgments}
We would like to thank Richard Kadel for motivating this study, as
well as Anna Hayes for assistance and for providing tabulated nuclear
data.  Critical discussions with Karsten Heeger, David Jaffe, and Petr
Vogel helped elucidate this work.  Brian Fujikawa and Herb Steiner
gave very helpful suggestions during the preparation of this
manuscript.  This work was supported under DOE OHEP DE-AC02-05CH11231
and DE-FG02-14ER42064.
\end{acknowledgments}

\bibliographystyle{apsrev4-1}
\bibliography{neutrino.bib}

\end{document}